\documentclass{article}
\usepackage{amsmath,amssymb}
\usepackage{graphicx}

\begin{document}

\title{Glass Formation and Crystallization \\ of a Simple Monatomic Liquid}

\author{Tomoko Mizuguchi and Takashi Odagaki \\ Department of Physics, Kyushu University, Fukuoka 812-8581, Japan}

\maketitle

\begin{abstract}
A simple monatomic system in two dimensions with a double-well interaction potential is investigated in a wide range of temperature by molecular dynamics simulation. The system is melted and equilibrated well above the melting temperature, and then it is quenched to a temperature 88\% below the melting temperature $T_m$ at several cooling rates to produce an amorphous state. Various thermodynamic quantities are measured as a function of temperature while the system is heated at a constant rate. The glass transiton is observed by a sudden increase of the energy and $T_g$ is shown to be an increasing function of the cooling rate in the preparation process of the amorphous state. In a relatively-high temperature region, the system gradually transforms into crystals, and the time-temperature-transformation(TTT) curve shows a typical nose shape. It is found that the transformation time to a crystalline state is the shortest at a temperature
$14\sim 15\%$ below the melting temperature $T_m$ and that at sufficiently low temperatures
the transformation time is much longer than the available CPU time. This indicates that a long-lived glassy state is realized.
\end{abstract}

\section{Introduction}

For a long time, a lot of studies have been made in understanding the glass transition by experimental, theoretical and computational methods\cite{rev_exp, rev_theo, rev_com}. One of the difficulties in understanding the glass transition arises from the fact that almost all glass forming materials consist of many constituents. Even in a computer simulation, more than two components are needed in order to avoid crystallization\cite{kob_andersen,soft_core}. The problem in a simple monatomic system is that crystallization occurs in a relatively short time. Since a long time simulation is needed to investigate the slow dynamics around the glass transition point, a binary mixture is usually investigated.

In order to avoid the crystallization of simple liquids in simulations, the Dzugutov potential was proposed \cite{dzug1}. The Dzugutov potential has an additional maximum to the Lennard-Jones potential at a range typical of the next-nearest-neighbor coordination distance in close-packed crystals, which suppresses the crystallization. Indeed, the system keeps a metastable supercooled state for a long time and shows some characteristic behaviors of glass-forming liquids\cite{dzug1,dzug2}. However, it turned out to form a dodecagonal quasicrystal by freezing in a subsequent simulation\cite{dzug3}. 

In fact, features of a glass formation are incompletely understood even for the simplest system, so it is attractive to carry out the study using a sophisticated monatomic system. In multicomponent systems, the relaxation involves both the topological and the chemical ordering. A spherically interacting one-component model system provides an opportunity for separating these contributions and enables us to explore the topological effect to the relaxation dynamics around the glass transition point. It is useful for comparison with theories, since most theoretical works\cite{mct,fel,replica} assume a simple system.

For the purpose of studying the glass formation of a simple liquid, we employ a monatomic model system in two dimensions with a double-well interaction potential, the Lennard-Jones-Gauss(LJG) potential\cite{engel1, engel2}. In a previous report\cite{mizuguchi}, we showed that the LJG system can be vitrified in two dimensions and the glassy state at low temperatures is stable for a fairly long time in spite of a simple monatomic potential. In addition, for 3D systems the glass-forming ability of the LJG system has been tested and discussed\cite{hoang}. 

In this paper, we show the further study of the glassy feature in this LJG system, in addition, we investigate the aging-induced crystallization for the monatomic glass-forming system with MD simulations. In general simulations, the model system is chosen to favour the clarification of dynamical relaxations under supercooling, while the crystallization is highly suppressed. Therefore, correct discriptions of the relaxation process of the glass-forming liquids are still not really understood, and we focus attention on this point. In Sec.2, we explain our model and methods of MD simulation in detail. We present the evidence of vitrification of this system in Sec.3 and we show the results of the crystallization process in Sec.4. Finally, we discuss the results in Sec. 5.

\section{Model Potential}

We consider a system of atoms which interact isotropically through the Lennard-Jones-Gauss (LJG) potential.
\begin{equation}
V(r)=\epsilon_0 \left\{ \left( \frac{r_0}{r} \right)^{12}- 2\left( \frac{r_0}{r} \right)^{6} \right\}-\epsilon \exp \left( -\frac{(r-r_G)^{2}}{2\sigma ^{2}} \right)
\end{equation}
The LJG potential consists of the LJ potential and a pocket represented by the Gaussian function. It is a double-well potential for most values of the parameters with the second well at position $r_G$, of depth $\epsilon$, and width $\sigma$. We note that this kind of effective potential is known to exist in matals. The general form of pair potentials in metals consists of a strongly repulsive core plus a decaying oscillatory term \cite{goukin}. The LJG potential can be understood as such an oscillatory potential, cut off after the second minimum. This potential was proposed for the self-assembly of two-dimensional monatomic complex crystals and quasicrystals \cite{engel1}. It has also been used as a model of liquid water\cite{water} and as a potential to stabilize a given structure\cite{rechtsman}. We find that stable glassy states can be obtained by quenching simulation with several values for the parameters.

\begin{figure}
\centering
\includegraphics[width=0.6\linewidth]{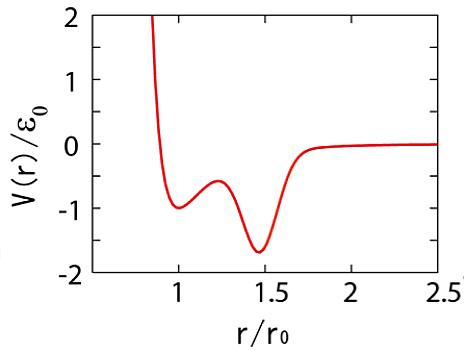}\\
\caption{The LJG potential for the parameters $r_G=1.47r_0, \epsilon=1.5\epsilon_0, \sigma^2=0.02r_0^2$}
\label{ljg}
\end{figure}

In this paper, we fix the parameter $r_G=1.47r_0, \epsilon=1.5\epsilon_0, \sigma^2=0.02r_0^2$. Figure \ref{ljg} shows the LJG potential with this chosen parameters. This set of parameters produces the pentagon-triangle phase in the ground state \cite{engel2}. The unit of length, energy and time in the present simulation are $r_0$, $\epsilon_0$, $\tau \equiv ( mr_0^2/\epsilon_0)^{1/2}$, respectively.

We perform MD simulations for 2,500 atoms on two dimensional space with periodic boundary conditions. The size of the simulation cell is chosen so that the number density becomes $\rho=1.0$. We employ a Verlet algorithm with time step $0.01\tau$.

\section{\label{sec:level1}Results and Discussions}
\subsection{\label{sec:level2}Glass Transition}
In order to determine the glass transition point, we follow the procedure employed in the thermodynamic measurement \cite{yamamuro}. Namely we prepare an amorphous state at a far below $T_m$ with a rapid cooling, and anneal the system for some time. Then we heat the system at a constant rate and measure the energy of the system. The glass transition temperature is identified by a change in the slope of E vs T graph.

\begin{figure}
\centering
\includegraphics[width=0.7\linewidth]{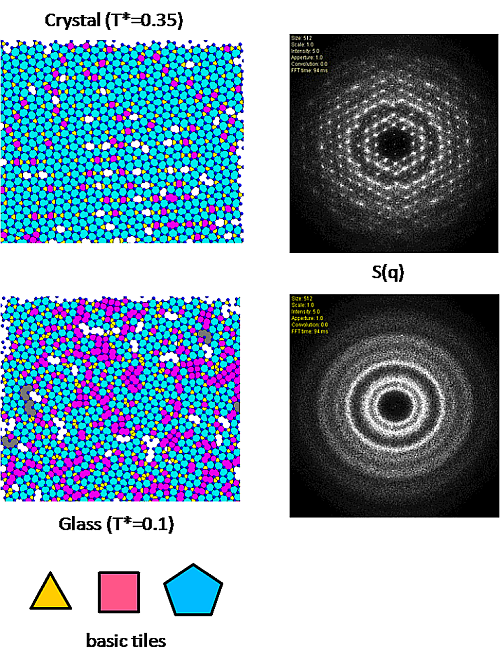}\\
\caption{Tiling structures in real space and diffraction images of a glass and crystal after a long run. A blue and yellow tile show pentagon and triangle, respectively. Square and hexagon are painted by same color, red. A gray tile shows collapsed nonagon.}
\label{structure}
\end{figure}

We prepare an equilibrium liquid state as the initial condition at $T^{*}\equiv k_{B}T/\epsilon_{0}=2.2$ which is above $T_m=0.43$ and then cool the system down to a target temperature with the velocity scaling. When the target temperature is close but below $T_m$, the system transforms into a crystalline state after some time. However, the target temperature is well below $T_m$, the system remains in the amorphous state even for a long time run. Figure \ref{structure} shows tiling structures in real space and diffraction images of a glass and crystal after $3.5\times 10^5\tau$. We determined tilings by drawing a line between the particles within a cutoff radius, where the cutoff radius is chosen as the first minimum of the pair correlation function of the system. Triangle and Pentagon tiles, which are painted by yellow and blue respectively, are essential for stable structures of this system. In the glassy state, any long range order cannot be observed.

\begin{figure}
\centering
\includegraphics[width=0.6\linewidth]{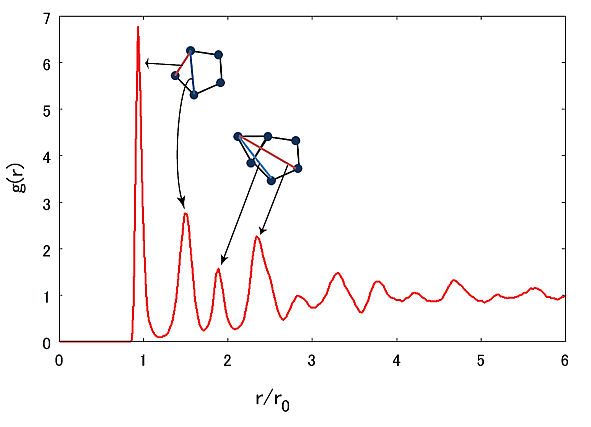}\\
\caption{The radial distribution function of a glassy state at $T^{*}=0.1$ and local structures.}
\label{gr}
\end{figure}

Figure \ref{gr} shows the radial distribution function of a glassy state. The posiiton of the first peak is 0.93 and the second peak is 1.50. The position of the second peak approximately coincides with the second minimum of LJG potential. The position of the first peak corresponds to a slightly inside from the first minimum of the potential. Here, we consider a regular pentagon, and if the distance between second nearest particles has the value of 1.47, the distance between nearest particles becomes about 0.91. This means the potential used in this study favors a pentagonal local order especially because of the position of the second minimum. This pentagonal local order is essential for the stability of the glass in this system. The third and fourth peak represent the existence of the local structure consisting of one pentagon and one triangle(see Fig. \ref{gr}). These tiles share on edge and make an ordered arrangement in a crystalline state, while it randomly spreads out all over the system in a glassy state(see Fig. \ref{structure}).

\begin{figure}
\centering
\includegraphics[width=0.6\linewidth]{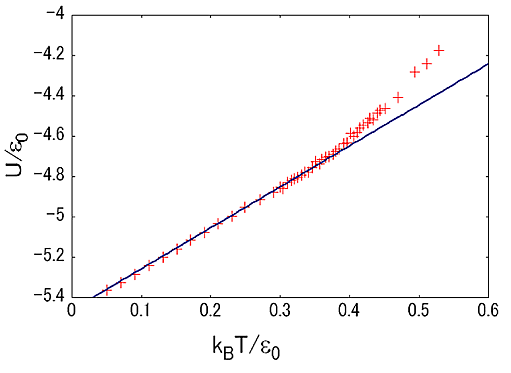}\\
\caption{The temperature dependence of the total energy in a heating process from a glassy state. If the system can be described only by the term of lattice vibration, plots are supposed to lie on the line in above figure.}
\label{Tg}
\end{figure}

\begin{figure}
\centering
\includegraphics[width=0.6\linewidth]{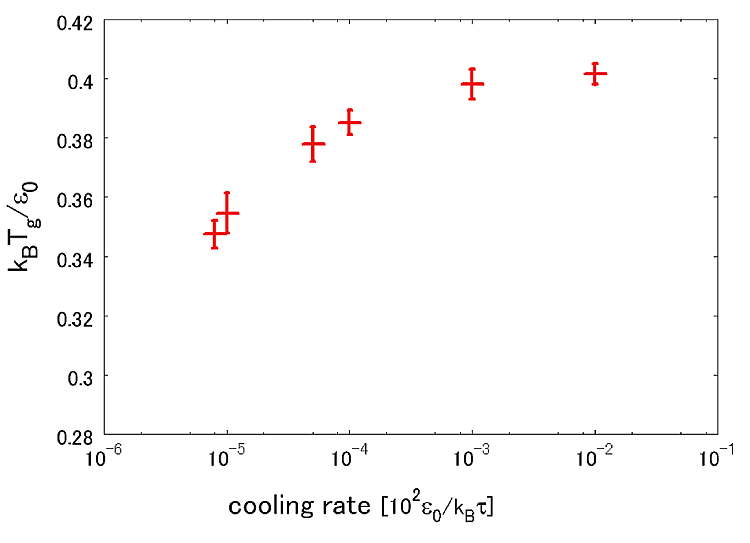}\\
\caption{The cooling rate dependence of $T_g$}
\label{cooling_rate}
\end{figure}

Figure \ref{Tg} shows the change of the total energy in a heating process from a glassy state at $T^{*}=0.05$ prepared with cooling rate $\gamma_q=10^{-2}\epsilon_0/k_B\tau$. The heating rate from a glassy state is $\gamma_h=10^{-3}\epsilon_0/k_B\tau$. The slope of the black line in Figure \ref{Tg} is about 2, which represents the energy change due to the vibration. Simulation data lie on the black line at lower temperatures, which indicates that the specific heat is determined by the vibrational motion of atoms. However, plots start to deviate from the line at $T^{*}\sim 0.38$, which is below the melting temperature, so that an additional effect appears in addition to that of vibration. Crystallization does not occur in this heating process, which is confirmed by the static structure factor. That means the system transforms from a glassy state into a supercooled state as the temperature is increased. If we take $\gamma_h=10^{-4}\epsilon_0/k_B\tau$ as the heating rate, the sign of crystallization appears in the heating process, so we cannot decide the glass transition point with confidence. Thus, we conclude the glass transition temperature is $T^{*}\sim 0.38$ in this sample. We perform the heating simulation starting from glassy samples prepared by different cooling rates and determine $T_g$ for each sample. Figure \ref{cooling_rate} shows the cooling rate dependence of $T_g$. We find that $T_g$ is an increasing function of the cooling rate, which is consistent with experiments \cite{rate_dep}.

\subsection{\label{sec:level2}Time-Temperature-Transformation Diagram}
In this section, we investigate the structural transformation into a crystal in the real
space. We first prepared an equilibrium liquid state at $T^* = 0.4$ above the melting
temperature $T^*_m\simeq 0.34$. Then, we quenched
the system to zero temperature instantaneously by removing the kinetic energy. We employ the instantaneous cooling in order to avoid any nucleation during the cooling process which may hinder proper comparison of the crystallization process.
After the quenching, we subjected the system to a heat reservoir at the target temperature and
allowed it to relax at this temperature.
The simulation cell is
made large enough so that the system feels zero pressure at all times. Under this free boundary condition, particles on the surface can easily move, so rearrangement of particles can frequently occur, compared with periodic boundary conditions. Since crystallization proceeds more drastically, we can determine easily the time when the system crystallizes. The number of atoms in the system is $N=1,024$ and the Leap-Frog algorithm is employed.

\begin{figure*}
\centering
\includegraphics[width=0.5\linewidth]{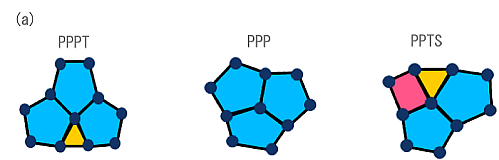}

\includegraphics[width=0.6\linewidth]{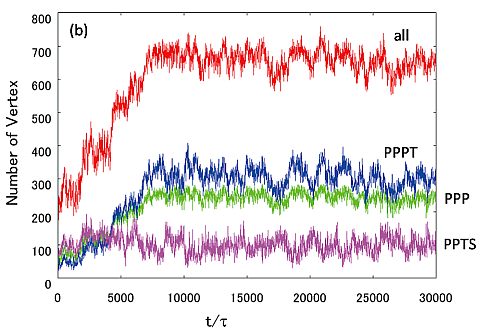}
\caption{(a) Three charateristic tiling structures.
(b) The time dependence of the number of characteristic tiling
 structures. The curve (all) represents the number of atoms surrounded 
by any of the pentagon, triangle or square tiles. The curves (PPPT), (PPP) 
and (PPTS) correspond to the structures shown in (a).}
\label{tiling}
\end{figure*}

We checked the time needed for crystallization at each temperature and determined a time-temperature-transformation(TTT) diagram. In order to determine the crystallization time,
we focus on the formation of three characteristic tiling structures,
PPPT, PPP and PPTS, which consist of pentagons, squares and
triangles shown in Fig. \ref{tiling}(a).
Figure \ref{tiling}(b) shows the time dependence of the number of
atoms whose surroundings can be identified as the three characteristic 
tiling structures.
The curve (all) shows the number of particles 
surrounded by any of the pentagons, triangles or squares. 
This value is related to the potential energy and becomes 
constant when the system reaches a crystalline state.
At relatively high temperatures, the transition from the supercooled liquid
to a crystal is rather sharp and we can determine the crystallization time
without ambiguity. As the temperature is reduced, 
the transformation tends to be less sharp, 
and we determined the transition time by observing diffraction patterns as well as
the evolution of tilings.

\begin{figure}
\centering
\includegraphics[width=0.6\linewidth]{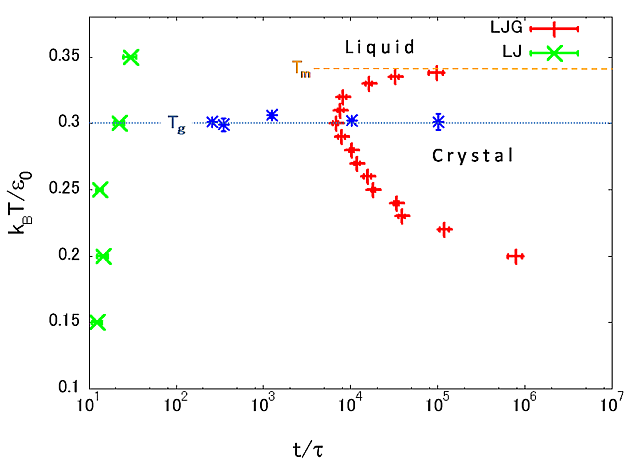}\\
\caption{The time-temperature-transformation diagram for monatomic LJ and LJG liquids}
\label{TTTdiagram}
\end{figure}

Figure \ref{TTTdiagram} shows the crystallization time as a function of temperature
which is known as a time-temperature-transformation (TTT) diagram.
This diagram indicates that the crystallization time becomes the shortest
at $\sim 0.86 T_m$ and it is longer than $10^7 \tau$ at $0.5 T_m$.
If one uses the value of $m$, $r_0$ and $\epsilon_0$ relevant for the LJ potential of Ar,
this time, $10^7 \tau$ corresponds to $10$ micro seconds. The glass transition temperature $T_g$ shown in Figure \ref{TTTdiagram} is determined under the following condition. First, we prepared glassy samples by instantaneous cooling to $0.15T_m$. After that, these samples were annealed at this temperature for $10\tau, 10^2\tau, 10^3\tau, 10^4\tau, 10^5\tau$ and we used these 5 samples as initial states for observation of glass transition. The temperature was increased from these initial states by $10^{-3}\epsilon_0/k_{B}$ a step and the energy was determined by its average for $1\tau$ period at each temperature step. Finally, we determined $T_g$ by a sudden increase in E vs T graph. We found that $T_g$ does not depend on the waiting time and is located at the temperature which a crystallization time is the shortest. 

For comparison, the results for a monatomic Lennard-Jones(LJ) system are also plotted
in Fig. \ref{TTTdiagram}. Consequently, the LJG system has a much longer crystallization time than the LJ system, and we can 
clearly see the temperature dependence of the TTT diagram in the LJG system. 
It shows a typical nose shape, which has been observed by experiments
for various glass forming materials\cite{mrs,TTT1,TTT2}. A similar TTT diagram is also found by MD simulations with some empirical potentials for metal\cite{TTT_Fe,TTT_Ni}. Our system has a much longer crystallization time than that reported in these papers in spite of the simpler shape of our interatomic potential.

A LJ liquid rapidly transforms into the triangle lattice at a few $10^3$ MD steps.  At low temperatures, particles condense into some clusters which are separated by some voids, and then such voids are eliminated gradually. On the other hand, at high temperatures, the system can transform in such a way that voids do not appear because particles can move more freely. This is the reason that the crystallization time in a LJ system becomes slightly longer at higher temperatures. 

\begin{figure*}
\centering
\includegraphics[width=0.6\linewidth]{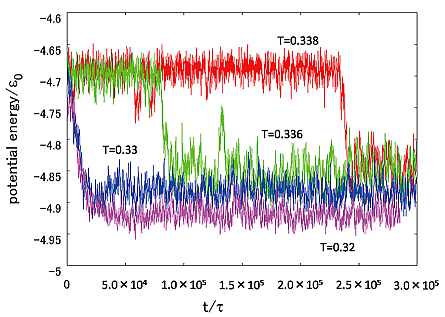}

\includegraphics[width=0.6\linewidth]{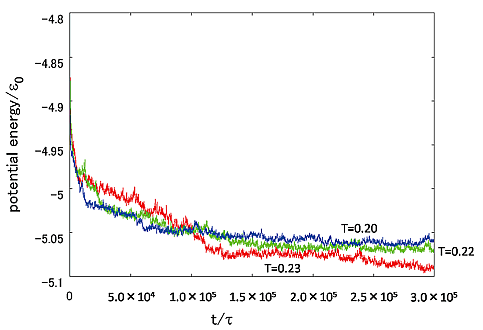}
\caption{Time dependence of potential energy at several temperatures. (a) shows the region close to $T_m$, 0.32$\leq T \leq $0.338. (b) shows the region at low temperatures, 0.20$\leq T \leq $0.23}
\label{energy}
\end{figure*}

\begin{figure*}
\centering
\includegraphics[width=0.35\linewidth]{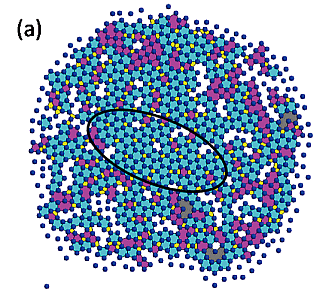}
\includegraphics[width=0.35\linewidth]{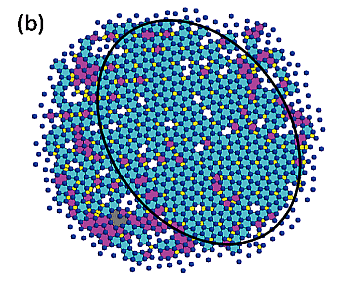}

\includegraphics[width=0.35\linewidth]{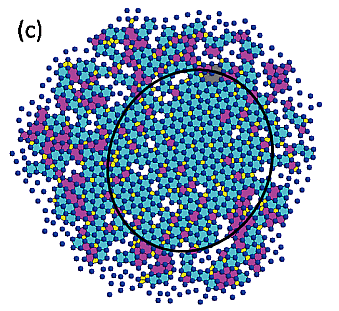}
\includegraphics[width=0.35\linewidth]{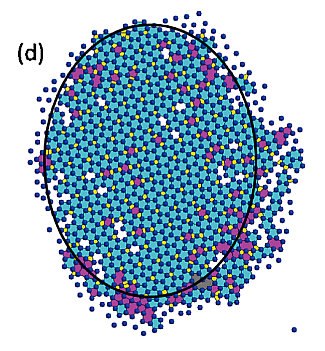}
\caption{The process of nucleation and growth at a temperature close to $T_m$. (a)The nucleation occurs in the center and (b) it grows rapidly. (c)A part of crystalline phase is desroyed and rearrangement occurs. (d)Eventually, the system becomes an almost perfect crystal.}
\label{nucleation}
\end{figure*}

Figure \ref{energy} shows results for the relaxation time dependence of potential energy at temperatures from 0.338 to 0.32 and from 0.23 to 0.20. It can be found from Figure \ref{energy}(a) that the system at temperatures very close to $T_m(=0.34)$ behaves as if it is an equilibrium liquid state for a while, and it drastically transforms into a crystalline state by random nucleation. The crystallization time in this temperature region is distributed because it depends on whether nucleation occurs or not by chance. Figure \ref{nucleation} shows one example of the process of nucleation and growth at temperatures close to $T_m$. The nucleation occurs in the inner part(Fig. \ref{nucleation}a) and grows rapidly(Fig. \ref{nucleation}b), however in a while, the order is destroyed from the outer side(Fig. \ref{nucleation}c). Since particles, especially on the surface, can move around rather freely, rearrangement can easily occur. Therefore, the system can remove defects and eventually reach a perfect crystal(Fig. \ref{nucleation}d). In Figure \ref{energy}b, the potential energy at a low temperature shows a higher value for long time scales compared with those at higher temperatures. It indicates that the system at sufficiently low temperatures keeps a metastable state for a long time. 

\begin{figure}
\centering
\includegraphics[width=0.6\linewidth]{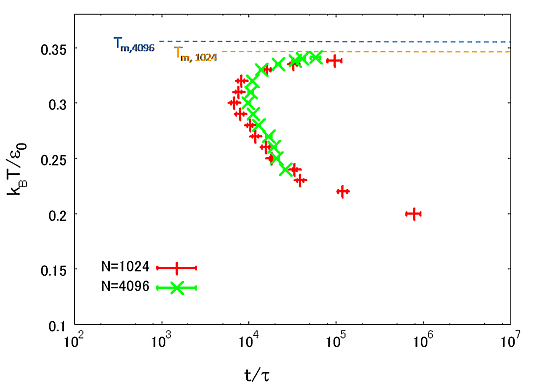}\\
\caption{The system-size dependence of the time-temperature-transformation diagram for monatomic LJG liquids. The number of particles in each system is 1024 and 4096. The temperature range in N=4096 system is from 0.24 to 0.341.}
\label{size_dep}
\end{figure}

Figure \ref{size_dep} shows the system-size dependence of TTT diagram. The number of particles in each system is $N=$1024 and 4096. In both systems, a longer time is needed for crystallization near $T_m$, and the temperature which has the shortest crystallization time is almost same. $T_m$ in a larger system is higher than that in a smaller system. This tendency agrees with the size dependence of $T_m$ found in confined systems\cite{mckenna}. We also find that the shortest time in a large system becomes longer than that of small systems. It is because, in a large system, a much longer time is needed for the spread of crystalline phase to the whole system. However, at lower temperatures, the growth of the crystallization time is suppressed in a large system. The reason of this result is that the state is different between two systems at the time when the system is assumed to have crystallized. At this time, many samples in the $N=$4096 system are a polycrystalline state which has more than two domains, on the other hand, the $N=$1024 system has less domains and sometimes one domain spreads to the whole system. If we determine the crystallization time when the system has one or two domains, a large system will take a longer time to crystallize.

\begin{figure}
\centering
\includegraphics[width=0.6\linewidth]{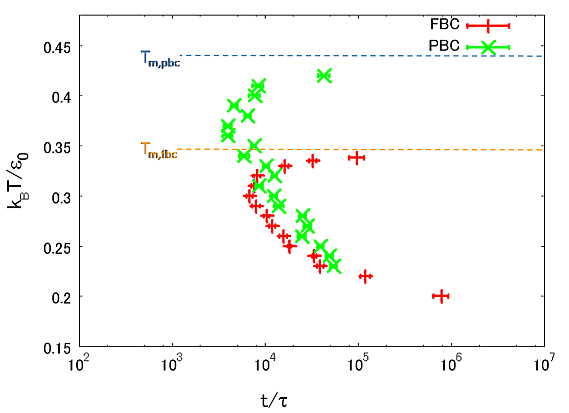}\\
\caption{The boundary condition dependence of the time-temperature-transformation diagram for monatomic LJG liquids}
\label{TTT_PBC}
\end{figure}

We also investigated the crystallization time under periodic boundary conditions(PBC). The boundary condition dependence of the TTT diagram is shown in Figure \ref{TTT_PBC}. The number of particles is 1024 in both systems. It is difficult to determine the crystallization time clearly under PBC in the same way as under free boundary conditions(FBC). In both systems, first nucleation occurs at almost same time. The difference appears in a subsequent process. In a system under FBC, if the nucleus starts to grow, a crystal phase spread rapidly to the entire system. It is more difficult for the nucleus to grow under PBC and a perfect crystal cannot be obtained unlike FBC. We determine the TTT diagram under PBC by drawing upon the potential energy. Since the index of crystallization is different, the crystallization time cannot be simply compared with one under FBC. 
The similarities between these two systems are that the crystallization time becomes longer at temperatures close to $T_m$ and it is the shortest at a temperature 15\% below $T_m$. The difference appears in a behavior at low temperatures. The growth of the crystallization time in PBC system is suppressed than that of FBC system. Also in this case, the state is different between two systems at the time when the system is assumed to have crystallized. A crystalline state in PBC system is less perfect than that of FBC system. It is more difficult in PBC system to get a perfect crystal because of boundary constraint.

\section{Conclusions}

We presented strong evidences of vitrification of a simple monatomic liquid in two dimensions by an adequate choice of LJG parameters. We have shown that the glassy state can survive a fairly long time at low
temperatures and the glass transition temperature is an increasing function of the cooling rate in the preparation process of the amorphous state. The Gaussian part of the LJG potential stabilizes
the pentagonal configuration and packing of pentagons produces
frustration in crystallization. Two competing length scales and the local structure made by them can have a great impact on the degree of disorder in the system, which leads to suppress crystallization. It has also been shown that, in 3D systems\cite{hoang}, the glassy state can be formed if the LJG potential has appropriate parameters favoring the formation of an icosahedral local order. The effect of this frustration with packing is the origin of the stability of the glassy state.

One of the features of this system is that the time needed for crystallization is sufficiently long. It enabled us to determine the TTT diagram of the simple system. It has a nose shape and the transformation time into crystal is the shortest at a temperature $14\sim 15\%$ below $T_m$ , which are independent of the system size and boundary conditions. Above the glass transition temperature, crystal nuclei rapidly spread to the whole system. Below $T_g$, the growth of crystal nuclei becomes slower as a temperature is decreased, because rearrangement of atoms occurs less frequent.

This model will enable us to explore in greater detail the mechanism of nucleation and growth processes in the monatomic glass. The study in this direction will lead us to solve the problem about slow nucleation kinetics of glass forming liquids. Due to simplicity of a  model, one can use the LJG system for further study of other phenomena related to the glass transition and for a detailed comparison with theories.

\medskip
\noindent
{\bf Acknowledgement}\\
We would like to thank Dr. Michael Engel of Stuttgart University
for providing us the software on which most of the simulations
and analysis for this work were carried out.
This work was supported in part by the Grant-in-Aid for Scientific Research
from the Ministry of Education, Culture, Sports, Science and Technology.

\end{document}